\documentclass[%
 reprint,superscriptaddress
]{revtex4-2}

\usepackage{graphicx}
\usepackage{dcolumn}
\usepackage{bm}
\usepackage{amsmath}
\usepackage{xcolor} 
\usepackage{hyperref} 
\usepackage[caption=false]{subfig}  
\usepackage{comment}   

\begin{document}


\title{Experimental Demonstration of Back-Linked Fabry–Perot Interferometer \\ for the Space Gravitational Wave Antenna}

\author{Ryosuke Sugimoto}
\email{rsugimoto@ac.jaxa.jp}
\affiliation{Graduate Institute for Advanced Studies, SOKENDAI, Sagamihara City, Kanagawa 252-5210, Japan}
\affiliation{Institute of Space and Astronautical Science, Japan Aerospace Exploration Agency, Sagamihara, Kanagawa 252-5210, Japan}

\author{Yusuke Okuma}
\affiliation{Department of Physics, University of Tokyo, Bunkyo, Tokyo 113-0033, Japan}
\affiliation{Institute of Space and Astronautical Science, Japan Aerospace Exploration Agency, Sagamihara, Kanagawa 252-5210, Japan}

\author{Koji Nagano}
\affiliation{Institute of Space and Astronautical Science, Japan Aerospace Exploration Agency, Sagamihara, Kanagawa 252-5210, Japan}

\author{Kentaro Komori}
\affiliation{Research Center for the Early Universe, University of Tokyo, Bunkyo, Tokyo 113-0033, Japan}
\affiliation{Department of Physics, University of Tokyo, Bunkyo, Tokyo 113-0033, Japan}

\author{Kiwamu Izumi}
\affiliation{Institute of Space and Astronautical Science, Japan Aerospace Exploration Agency, Sagamihara, Kanagawa 252-5210, Japan}

\date{\today}
\begin{abstract}
The back-linked Fabry-Perot interferometer (BLFPI) is an interferometer topology proposed for space gravitational wave antennas with the use of inter-satellite Fabry-Perot interferometers. The BLFPI offers simultaneous and independent control over all interferometer length degrees of freedom by controlling the laser frequencies. Therefore, BLFPI does not require an active control system for the physical lengths of the inter-satellite Fabry-Perot interferometers. To achieve a high sensitivity, the implementation must rely on an offline signal process for subtracting laser frequency noises. However, the subtraction has not been experimentally verified to date. This paper reports a demonstration of the frequency noise subtraction in the frequency band of 100~Hz-50~kHz, including the cavity pole frequency, using Fabry-Perot cavities with a length of 46~cm. The highest reduction ratio of approximately 200 was achieved. This marks the first experimental verification of the critical function in the BLFPI.
\end{abstract}

\maketitle


\section{\label{sec:level1}INTRODUCTION}
Since the groundbreaking observation of gravitational waves radiated from a binary black hole in 2015~\cite{LIGO150914}, the network of terrestrial gravitational wave interferometers finished conducting three observing runs in March 2020, reporting  the observations of 90 gravitational wave event candidates from compact binary coalescences~\cite{GWTC-3}. Currently, the terrestrial network is in the fourth observing run. All the events so far were identified to be those radiated from binary systems consisting of either two neutron stars, two stellar mass blacks hole or the combination of the two. They are found to be in the mass range of $1-100\,{M_\odot}$, corresponding to the observation frequency band of 10~Hz-1~kHz~\cite{obsscenario}. In June 2023, the North American Nanohertz Observatory for Gravitational Wave, one of the pulsar timing array (PTA) experiments, reported the observational evidence of detection of a nHz gravitational wave background, likely from the ensemble of supermassive black hole binaries~\cite{Agazie2023}.

On the other hand, the frequency band of $\rm{mHz}-10\ \rm{Hz}$ remains unexplored as a frequency gap, given the fact that the frequency band of terrestrial interferometers is limited by ground vibration noises~\cite{Rana2014} and that of the PTA experiments is limited by the integration time. Complementing this gap is of high importance because it would offer observations of the binary systems of new mass range~\cite{Matsubayashi2004, Reisswig2013}, and cosmological sources~\cite{Grishchuk1975, kuroyanagi2009}. Several strategies have been proposed to fill the frequency gap. They include plans to utilize torsional pendulums with low resonant frequency for effective vibration isolation, as represented by TOBA~\cite{TOBA}; atomic interferometers that employ free-fall atoms to mitigate disturbances, such as AION~\cite{AION}, ZAIGA~\cite{ZAIGA}, and AEDGE~\cite{AEDGE}; plans to utilize the stable cryogenic materials to remove seismic noise with a high common mode rejection, exemplified by SOGRO~\cite{SOGRO}, as well as a number of space-borne interferometer missions proposals, that essentially avoid the terrestrial environment, such as LISA~\cite{LISA2017}, Taiji~\cite{Taiji2020}, TianQin~\cite{TianQin2021}, BBO~\cite{Harry_2006}, TianGO~\cite{PhysRevD.102.043001}, B-DECIGO, DECIGO and other concepts~\cite{ni2016gravitational}. 

In particular, B-DECIGO~\cite{preDECIGO2016}, an inter-satellite Fabry-Perot interferometer, has the potential to achieve a high sensitivity in the 0.1~Hz band. It is capable of observing the intermediate-mass black hole binaries with a total mass of 100-$10^4M_\odot$ up to a redshift of $\sim 300$ with a signal-to-noise ratio of 8. It will also be capable of detecting neutron star binaries before they merge. For instance, B-DECIGO should be able to detect such a system 7 years before the merger if it is at a distance of 40~Mpc~\cite{Isoyama2018} comparable to GW170817. DECIGO, the ambitious successor of B-DECIGO, will improve the sensitivity by an order of magnitude and is hoped to achieve a direct observation of primordial gravitational wave backgrounds~\cite{DECIGO}.

\begin{figure*}[htb]
\centering
\begin{tabular}{c}
\begin{minipage}[b]{0.45\linewidth}
\subfloat[\label{fig:BLFP1}]{%
  \includegraphics[keepaspectratio, scale=0.235]{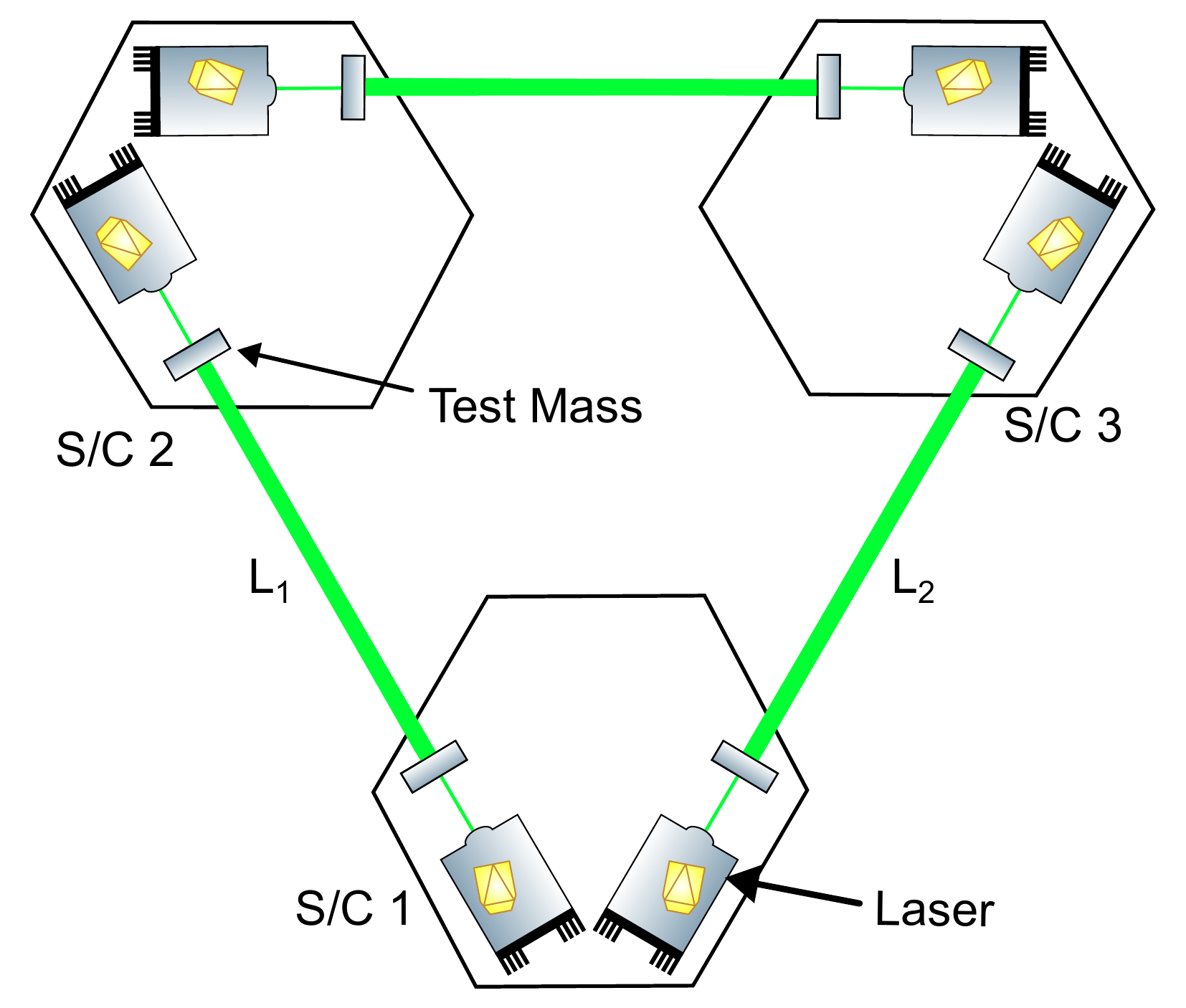}%
}
\end{minipage}\hfill

\begin{minipage}[b]{0.45\linewidth}
\subfloat[\label{fig:BLFP2}]{%
  \includegraphics[keepaspectratio, scale=0.355]{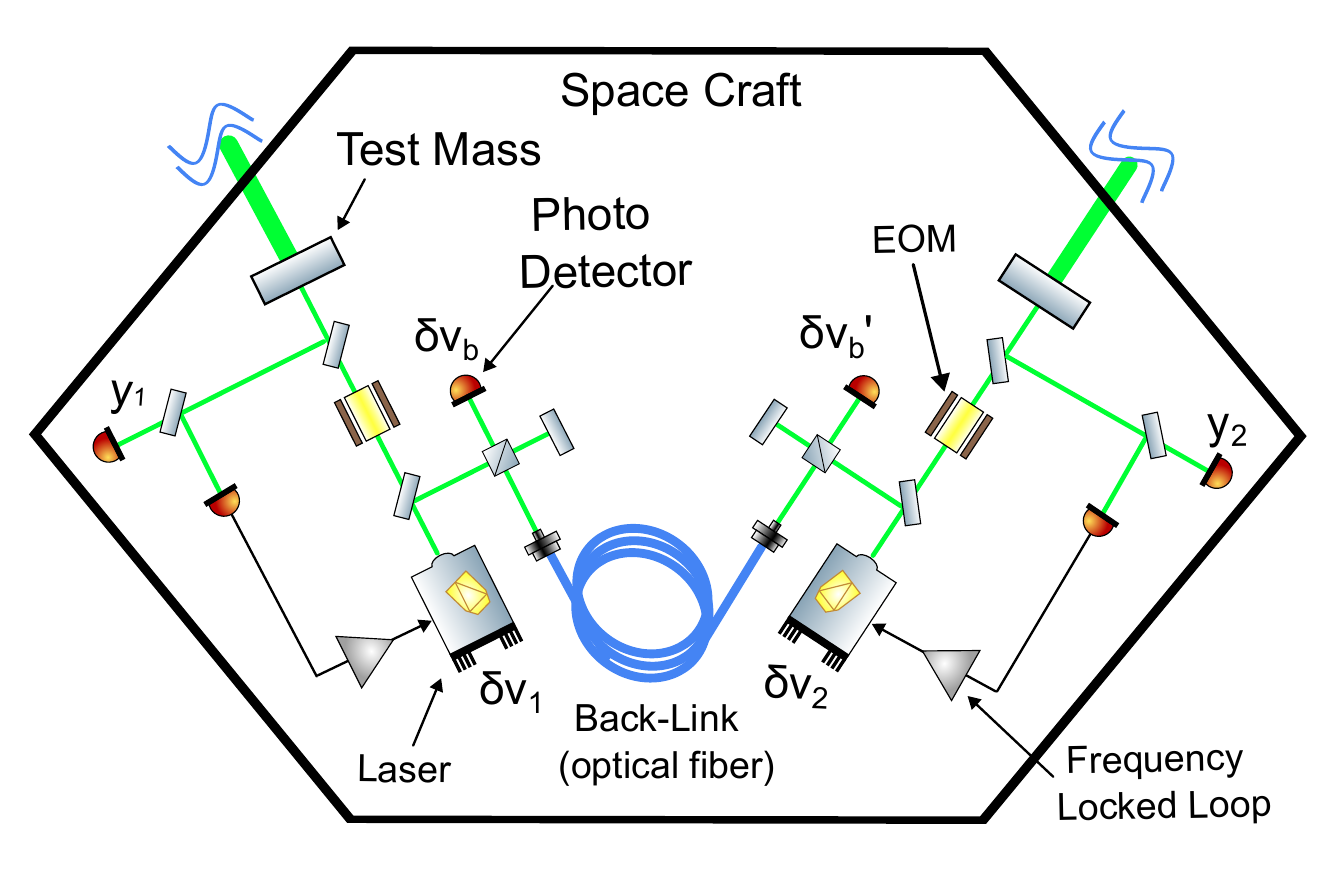}%
}
\end{minipage}
\end{tabular}
 \caption{(a) Schematic view of the BLFPI. Three inter-satellite interferometers are constructed by three S/C (spacecraft). Each spacecraft is equipped with two laser sources so that all lasers can be locked to the cavities by frequency locked loops without controlling the position of test masses. (b) Schematic of the internal optical system of each S/C. The heterodyne beat signals as $\delta \nu_\mathrm{b}$ and $\delta \nu_\mathrm{b}'$ are acquired at each end of the back-link path with two photodetectors. The combination of the two heterodyne signals can subtract phase noises introduced in the back-link fiber in the post-process. The reflected light from each cavity is monitored by two photodetectors, one for laser frequency control and the other out-of-loop photodetectors for obtaining the $y_1$ and $y_2$ signals for the subtraction.}\label{fig:BLFP}
\end{figure*}

The back-linked Fabry-Perot interferometer (BLFPI) was proposed~\cite{BLFP} as an interferometer topology using the inter-satellite Fabry-Perot interferometers (see Fig.~\ref{fig:BLFP}). We colloquially call the Fabry-Perot interferometer the Fabry-Perot cavity or cavity in short, hereafter. The BLFPI keeps all the cavities at a resonance by controlling the laser frequencies only, thus avoiding the need for precision control of the inter-satellite distances at an unprecedented precision level of nanometers. Such a precision control would be required if the resonances were maintained by controlling the physical lengths of cavities. The BLFPI is expected to overcome the serious design problem where the amount of propellant stringently limits the length of the observation period due to continuous control of the satellite positions. In addition, if the inter-satellite cavity lengths were controlled for keeping the resonances, one cavity length degree of freedom would need to be left uncontrolled, requiring an additional adjustment for the cavity lengths~\cite{Nagano_2021}. Since the BLFPI employs a simple control configuration in which each laser is locked to each corresponding cavity, there is no restrictions on the values of cavity lengths, allowing the satellites to form an arbitrary and time-variant triangular formation. Finally, the simple configuration makes the in-orbit operations more tractable such as lock acquisition.

While the BLFPI offers several advantages, it is vulnerable to laser frequency noises which would significantly deteriorate the observatory sensitivity if unaddressed. For the reason, the BLFPI was proposed together with a new offline noise subtraction scheme similarly to time-delay interferometry~\cite{Tinto1999}. In the BLFPI, the heterodyne beatnote signals are obtained by optically connecting the two lasers with an optical fiber called the back-link similarly to LISA~\cite{Steier2009,Fleddermann2018,Isleif2018} in addition to the error signals from each cavity. The success of BLFPI heavily relies on this subtraction process which has not experimentally been tested to date. 

This paper for the first time reports an experimental demonstration of the BLFPI. A miniature of the BLFPI was built on an optical bench with the main aim of validating the frequency noise subtraction. We show that the frequency noise can be subtracted to a reduction ratio of $188 \pm 29$ and discuss the current limitations for the subtraction.

This paper is organized as follows. In Sect~\ref{sec:2}, we introduce the frequency noise subtraction scheme and show that the frequency noises can be subtracted while leaving the gravitational wave signals or equivalently the length signals unaffected. In Sect~\ref{sec:3}, the experimental setup and the implementation of the offline subtraction process are presented. In Sect~\ref{sec:4}, it is shown that the laser frequency noises were successfully reduced by adapting the subtraction process to the data acquired in the experiment. Sect~\ref{sec:5} discusses a number of sources that could limit the noise subtraction ratio and the implication to the space gravitational wave antennas. Finally, Sect~\ref{sec:6} summarizes the paper.

\section{frequency noise subtraction}\label{sec:2}
We briefly explain the frequency noise subtraction and set the definitions of parameters relevant to the experiment. While the proposal paper~\cite{BLFP} handles frequency noises in terms of phase in rad, we treat them in terms of frequencies in Hz such that they can be easily translated to the measurement values in the experiment. 

\begin{figure*}[t]
\includegraphics[keepaspectratio, width=132mm]{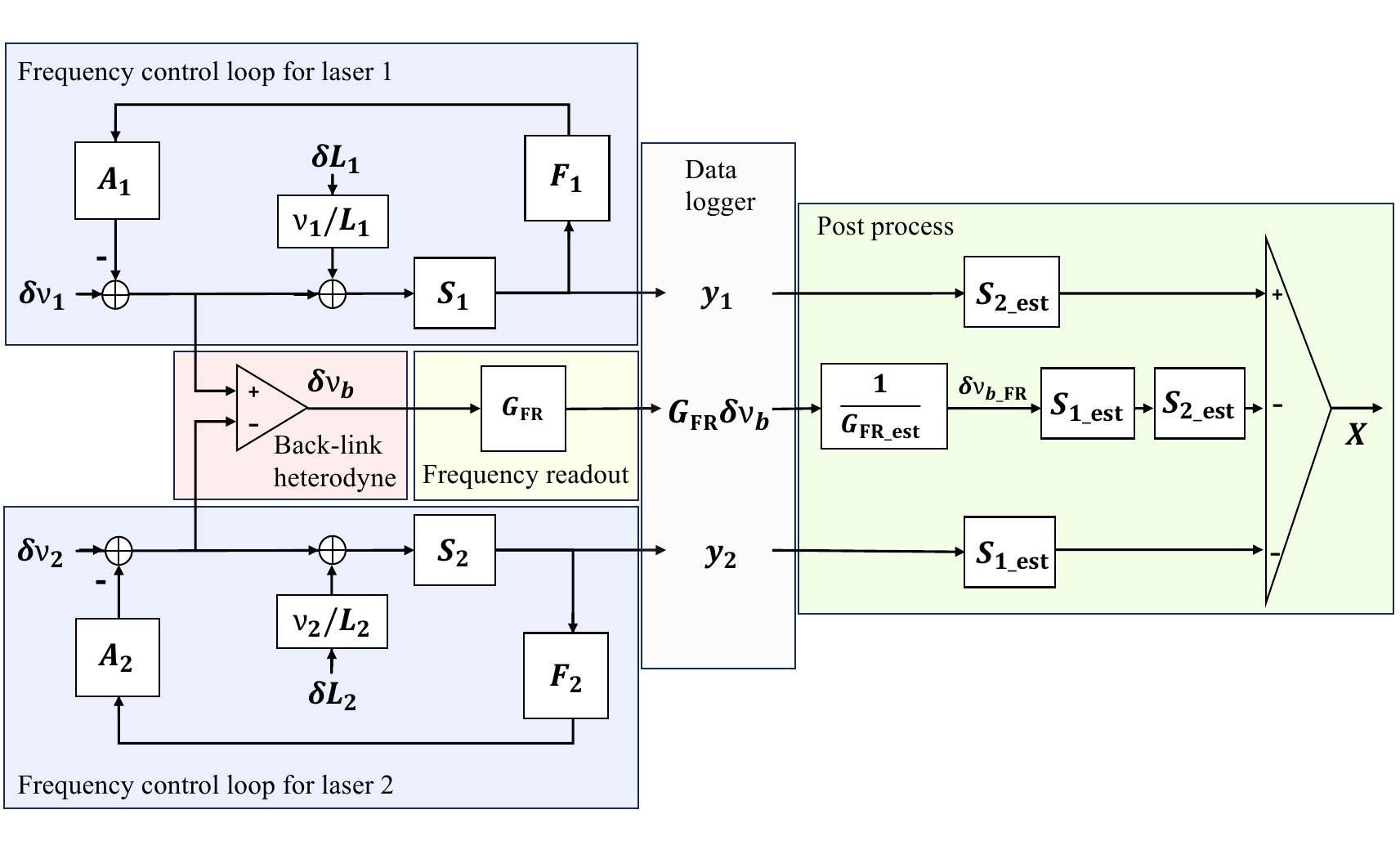}
\caption{\label{fig:block} Block diagram of the control system and post process for the BLFPI. The system consists of the experimental system for signal acquisition and the post process system to subtract laser frequency noises. The experimental system is further divided into the control system to lock the lasers to a cavity resonance, the back-link part to monitor the heterodyne beat signals, and the frequency readout part to monitor the time evolution of the beatnote frequency.}
\end{figure*}

Figure~\ref{fig:block} illustrates the block diagram for a single interferometer of the BLFPI constellation. The frequency of each laser source is locked to the corresponding cavity length via a liner servo filter $F_j$ with $j=1,\,2$. The frequency of each laser comes with fluctuations or frequency noise denoted by $\delta\nu_j$. The frequency locking servo then suppresses frequency noises by feeding the signal back to a frequency actuator $A_j$. Consequently, the laser frequency tracks the cavity length fluctuation $\delta L_j$.

The error signals for the $j$-th laser source with respect to the length of cavity can be expressed in the frequency domain as
\begin{equation}
y_j=\frac{S_j}{1+G_j} \delta \nu_j+\frac{S_j}{1+G_j} \frac{\nu_j}{L_j} \delta L_j,
\end{equation}
where $S_j$ and $G_j$ describe a sensing transfer function converting frequency into voltage and open loop gain defined by $G_j \equiv S_j F_j A_j$, respectively. 

The back-link path optically combines the two laser beams before they are incident on the Fabry-Perot cavities, producing heterodyne beatnote signals. Due to the frequency-locked loops, the beatnote signals contain the frequency noises of laser sources as well as the length fluctuation of the Fabry-Perot cavities. The beat signal in terms of its frequency shift $\delta \nu_\mathrm{b}$ can be expressed as
\begin{equation}
\begin{aligned}
\delta \nu_{\mathrm{b}}&=& \frac{1}{1+G_1} \delta \nu_1-\frac{G_1}{1+G_1} \frac{\nu_1}{L_1} \delta L_1\\
&-&\frac{1}{1+G_2} \delta \nu_2+\frac{G_2}{1+G_2} \frac{\nu_2}{L_2} \delta L_2.
\label{eq:one}
\end{aligned}
\end{equation}
The frequency shift can be read out by, for example, a phasemeter. The conversion factor from frequency to voltage is given by $G_\textrm{FR}$.

Once the three measurement data $y_j$ and $\nu_\mathrm{b}$ are recorded at the same time, one can obtain the synthesized signal $X$ as
\begin{equation}
\begin{aligned}
X&=\delta \nu_{b \textrm{\_FR}} S_{1 \textrm{\_est}} S_{2\textrm{\_est}}-\left(y_1 S_{2 \textrm{\_est}}-y_2 S_{1 \textrm{\_est}}\right) \\
& =\left[\frac { 1 } { 1 + G _ { 1 } } \left\{\left(\frac{G_\textrm{FR}}{G_\textrm{FR \_est}}-\frac{S_1}{S_{1 \textrm{\_est}}}\right) \delta \nu_1\right.\right. \\
& \left.-\left(\frac{G_\textrm{FR}}{G_\textrm{FR \_est}} G_1+\frac{S_1}{S_{1 \textrm{\_est}}}\right) \frac{\nu_1}{L_1} \delta L_1\right\} \\
& -\frac{1}{1+G_2}\left\{\left(\frac{G_\textrm{FR}}{G_\textrm{FR \_est}}-\frac{S_2}{S_{2 \textrm{\_est}}}\right) \delta \nu_2\right. \\
& \left.\left.-\left(\frac{G_\textrm{FR}}{G_\textrm{FR \_est}} G_2+\frac{S_2}{S_{2 \textrm{\_est}}}\right) \frac{\nu_2}{L_2} \delta L_2\right\}\right] S_{1 \textrm{\_est}} S_{2 \textrm{\_est}}\label{eq:3}. 
\end{aligned}
\end{equation}
The quantities with the subscript ``est'' represent the estimated parameters for the offline subtraction process.
Assuming that $S_1$, $S_2$, and $G_\textrm{FR}$ can be estimated with sufficiently high accuracy so that $S_{1 \_est}\rightarrow S_1$, $S_{2 \_est}\rightarrow S_2$, and $G_\textrm{FR \_est}\rightarrow G_\textrm{FR}$, one can arrive at
\begin{equation}
\begin{aligned}
X&  = \left(-\frac{\nu_1}{L_1} \delta L_1+\frac{\nu_2}{L_2} \delta L_2\right) S_1 S_2 \\ 
& =\left\{\left(-\frac{\nu_1}{L_1}+\frac{\nu_2}{L_2}\right) \delta L_{+}-\left(\frac{\nu_1}{L_1}+\frac{\nu_2}{L_2}\right) \delta L_{-}\right\} S_1 S_2 \\
& \approx\left\{\frac{-\Delta L}{L_1\left(L_1+\Delta L\right)} \delta L_{+}-\frac{2 L_1+\Delta L}{L_1\left(L_1+\Delta L\right)} \delta L_{-}\right\} S_1 S_2 \nu,
\end{aligned}
\end{equation}
with $\delta L_\pm$ being the common and differential components of two lengths defined by
\begin{equation}
    \delta L_{ \pm} \equiv \frac{\delta L_1 \pm \delta L_2}{2}.
\end{equation}
In addition, we have approximated the laser frequencies as $ \nu \approx  \nu_1 \approx  \nu_2$. This is valid as long as the beatnote frequency is approximately 100~MHz or smaller for the laser frequency of several 100 THz, as mentioned in Sect~\ref{sec:3}. In this case, the relative difference between the two laser frequencies should be less than 1~ppm. Two cavity lengths are related via $L_2 \equiv L_1 + \Delta L$ where $\Delta L$ is the macroscopic difference.

The synthesized output now contains only the displacement of the cavity length. In the event of gravitational waves passing through the Fabry-Perot interferometer, they appear as the differential mode of the lengths with the laser frequency noises subtracted. This is the working principle of the BLFPI and frequency noise subtraction. As is apparent from Eq.~(\ref{eq:3}), the success of frequency noise subtraction largely depends on the accuracy in the estimation of the interferometer responses $S_1$ and $S_2$. Additionally, the coefficients for the frequency readout $G_\textrm{FR}$ also affects the subtraction performance. However, it can be independently calibrated with an oscillator of known amplitude and frequency.

\section{Experimental setup}\label{sec:3}
\begin{figure*}[btp]
\includegraphics[keepaspectratio, width=\linewidth]{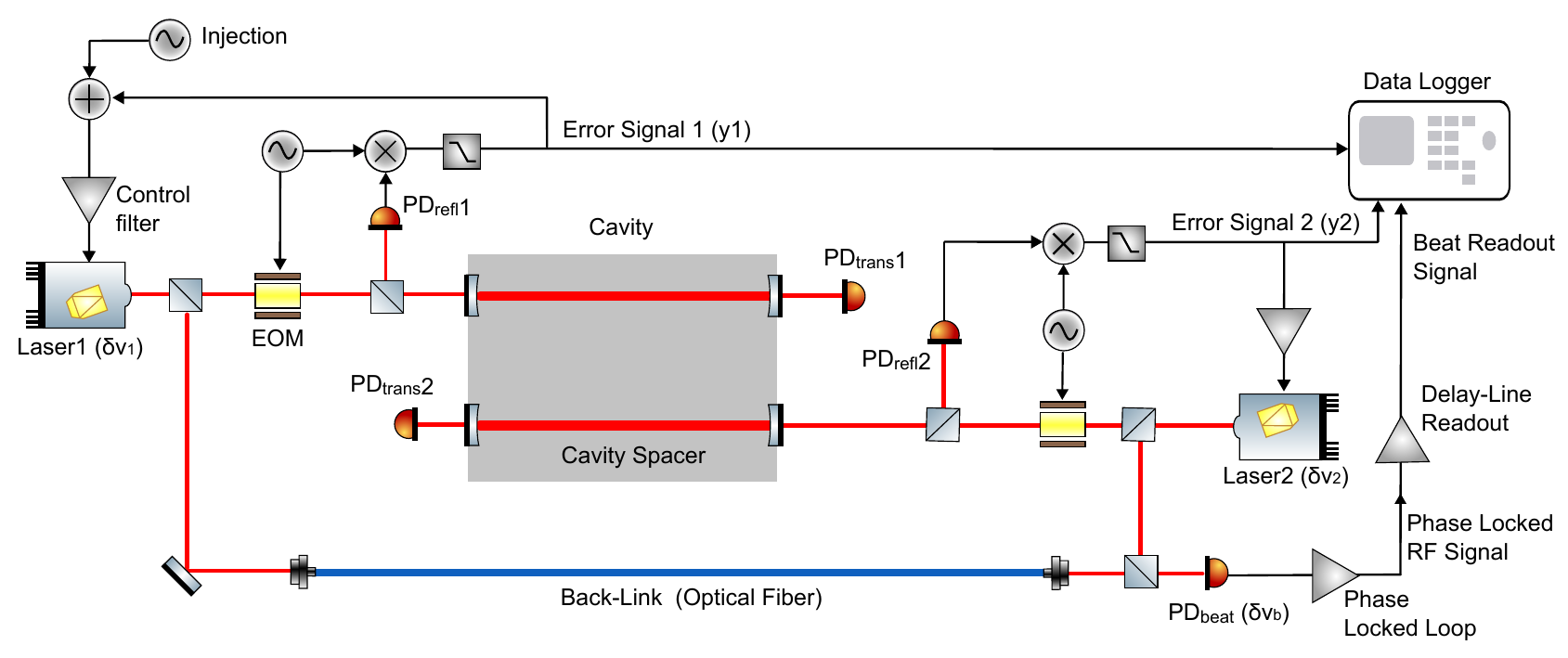}
\caption{\label{fig:layout} Experimental setup for demonstration. It consists of the laser frequency locked loops using the PDH readout and the back-link part for producing the heterodyne beatnote signals. The frequency fluctuation of the heterodyne beatnote signal is converted to voltage signals by the combination of the phase locked loop and delay line. The frequency fluctuations is then recorded by the data logger along with the error signals. The difference from the spacecraft layout shown in Fig.~\ref{fig:BLFP2} is that the error signal used for subtraction process is shared with the in-loop signal used for laser frequency lock. Since the phase noise expected to be introduced from the 1~m length backlink optical fiber~\cite{Enomoto_PhD} is almost negligible compared to the expected noise floor shown in Fig.~\ref{fig:beat_budget}, the heterodyne beat signal is acquired with only one photodetector. In addition, square waves to excite the frequency noise are injected for the demonstration at a point before the control filter of the frequency locked loop for laser 1.}
\end{figure*}

For the practical application of the BLFPI, it is necessary to conduct the experimental test to demonstrate the frequency noise subtraction and possibly identify issues associated with the implementation. The experimental setup for our demonstration is illustrated in Fig.~\ref{fig:layout} where an equivalent of the single BLFPI miniature is built on a tabletop under the atmospheric pressure. Two independent laser sources at a wavelength of 1064~nm are employed. Each laser field is divided into two paths by an unpolarized beam splitter with a splitting ratio of 50:50. One is the optical path to the cavity, and the laser field is phase-modulated by an Electro Optic Modulator (EOM) at a radio frequency before the Fabry-Perot cavity, allowing for the Pound-Drever-Hall (PDH) readout scheme~\cite{drever1983laser}. The other backlink optical path is implemented by a 1~m of optical fiber sending the light of a laser source to the other, producing the heterodyne beatnote. Each laser light has a power of about 5~mW before the beamsplitter.

The frequency of each laser is locked to the cavity length by feeding the PDH signals with a linear control filter. Therefore, two Fabry-Perot cavities are locked at resonance points.

The two Fabry-Perot cavities have an identical design length value of 46~cm. The finesse of cavity 1 was measured to be approximately 11000, and 14000 for cavity 2. We estimated the cavity pole frequency to be approximately 15~kHz and 12~kHz, for cavities 1 and 2, respectively based on the measurement values for finesse. The design choices were made primarily to have the cavity pole low enough such that the cavity pole feature can be captured by the data acquisition system at a realistic sampling rate. This is because, as shown in Table~\ref{tab:DECIGO}, detectors such as DECIGO include the cavity pole in the observation bands, and thus it is important to demonstrate the subtraction of frequency noise in the frequency bands~\cite{DECIGO} around the cavity pole. A dedicated spacer accommodating the two Fabry-Perot cavities is designed and implemented. The spacer is made of a super invar, IC-36FS, from Shinhokoku Material Corporation with a dimension of $80 \times 460\times 40$~mm$^3$. The thermal expansion coefficient is expected to be 0.8~ppm/K so that the effect from ambient temperature is small. The cavity mirrors are attached to the spacer with a combination of acrylic plate and rubber ring. Additionally, we installed an aperture mask in both intra-cavities to suppress the excitation of high-order spatial modes.
\begin{table}[b]
\caption{\label{tab:DECIGO}%
Comparison of the experimental setup and space detectors for the main parameters of the optical cavity.}

\begin{ruledtabular}
\begin{tabular}{lcc}
& \begin{tabular}{c} Experimental setup
\end{tabular}
 & DECIGO \\
\hline
\hline
\begin{tabular}{l} Cavity length [m]
\end{tabular}& $0.46$ & $1.0\times10^6$ \\\hline
\begin{tabular}{l} Free spectral range [Hz]
\end{tabular}& $\sim 3.3\times10^8$  & $ 1.5\times10^2$ \\\hline 
\begin{tabular}{l} Finesse
\end{tabular}& $\sim 10^4$ & $10$ \\\hline
\begin{tabular}{l} Cavity pole [Hz]
\end{tabular}& $\sim 10^4$ & $7.5$ \\\hline
\begin{tabular}{l} Frequency band [Hz]
\end{tabular}& $100- 5.0\times10^4$ & $0.1-10$ \\\hline
\begin{tabular}{l} Laser\\ wave length [nm]
\end{tabular}& $1064$ & $515$ \\\hline
\begin{tabular}{l} Cavity Incident\\ Laser Power [W]
\end{tabular}& $\sim 2.5\times10^{-3}$ & $10$ \\
\end{tabular}
\end{ruledtabular}
\end{table}

Two types of signals are recorded in the measurements. One is the error signals derived by the PDH scheme, representing the relative measure of laser frequency with respect to the cavity length. The setup contains two such signals, namely $y_1$ and $y_2$. The other type is the frequency variations of the beatnote generated by the back-link heterodyne. This beatnote signal is designed to produce a monochromatic oscillative signal at 10-160~MHz. The center frequency of the beatnote signal can change depending on the resonant conditions of the Fabry-Perot cavities. The beatnote signal is then fed to a commercial phasemeter which tracks the phase evolution via the digital phase-locked loop. The phasemeter then produces an analog copy of the input signal with a fixed amplitude to reduce the adverse effect from amplitude fluctuations and higher harmonics. Finally, the analog copy of the beatnote signal is converted into the voltage signal in proportional to the beatnote frequency via the delay line fequency discreminator~\cite{DLD1985}. Although the phasemeter has the function of recording the phase evolution, we did not use it. The use of the phase record function will require an additional system to synchronize the phasemeter and data logger.

\begin{figure}[btp]
\includegraphics[keepaspectratio, width=\linewidth]{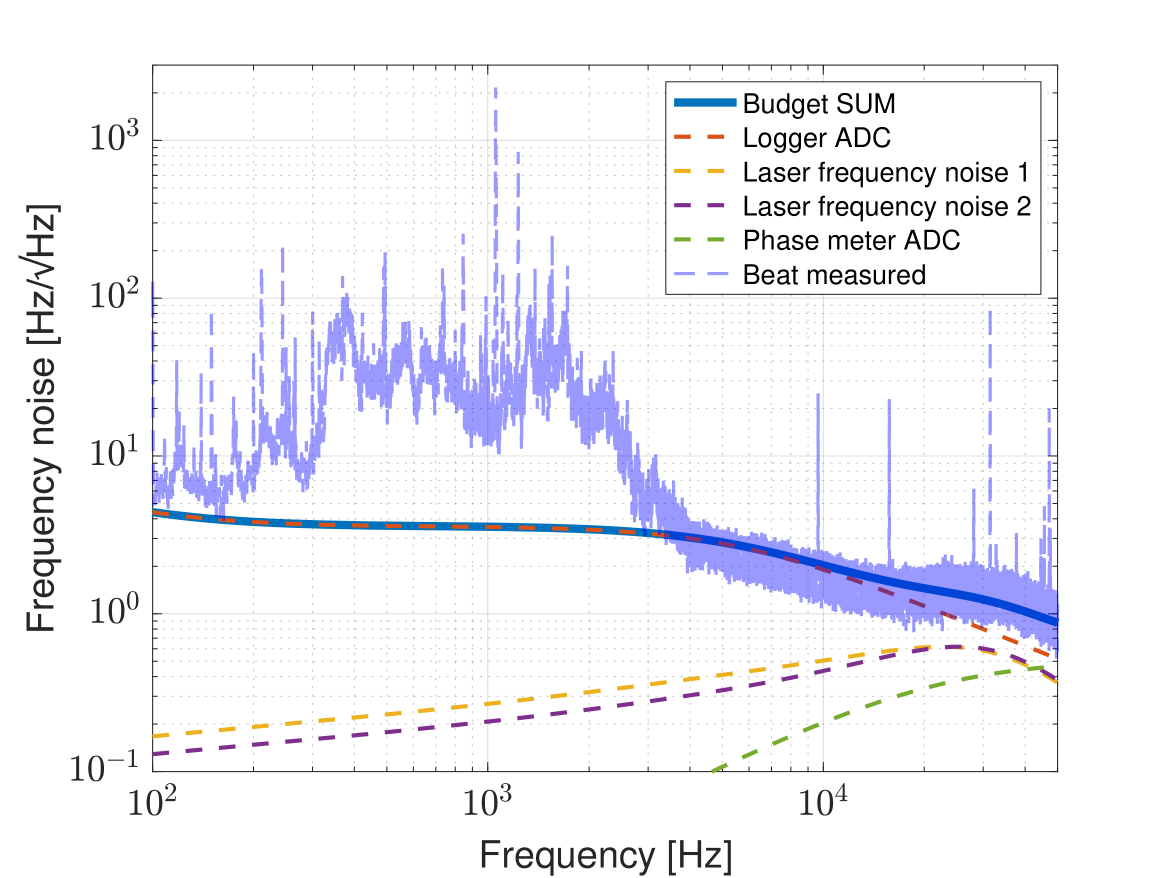}
\caption{\label{fig:beat_budget} Amplitude spectral densities of measured frequency fluctuation at the heterodyne beatnote $\delta \nu_\mathrm{b}$ and various estimated noise contributions. (Dashed blue) Measured noise without an excitation. (Thick blue curve) Quadrature sum of all known noise contributions. (Red dashed) digitization noise of ADC or analog-to-digital converter for the logger used for recording the heterodyne beatnote. (Dashed yellow) Suppressed laser frequency noise from laser 1. (Dashed purple) the same for laser 2. (Dashed green) ADC digitization noise for the digital phase-locked loop in the phasemeter.}
\end{figure}

The subtraction performance was evaluated by observing changes in the spectra referred to the beatnote frequency in Hz/Hz$^{1/2}$ with and without the subtraction applied. The spectrum of the beatnote frequency without the subtraction was determined to be dominated by technical noises as summarized in Fig.~\ref{fig:beat_budget}. Unexpected acoustic noise exists in a broad band up to several kHz. Since these acoustic noises are not present when the cavities are not resonant, they are likely be associated with the frequency locked loops. Possible noise sources include the residual of the differential cavity fluctuation and the sensing noise of the control. The ADC noise of the data logger, which is dominant at high frequencies, can be reduced by introducing a preamplifier if lower noise is required in future.

Since the laser frequency noises are far below the technical noises across the frequency band, a function generator was hooked up to laser 1 to deliberately increase the magnitude of the frequency noise. The square waves at a fundamental frequency of 113 Hz are applied to increase the laser frequency noise in a broad frequency range. The excitation can be switched off whenever necessary.

The subtraction process was implemented in the frequency domain as opposed to the time domain. As pointed out in time-delay interferometry for LISA~\cite{Tinto1999}, the frequency domain subtraction introduces undesired residuals due to the finite-length Fourier transform. Nonetheless, in our demonstration the subtraction ratio is not limited by that and therefore adopted the frequency domain treatment for simplicity.

The data streams $y_1$, $y_2$ and $\nu_{b \textrm{\_FR}}$ are divided into $2n-1$ data chunks each with a 50\% overlap so that Welch's method ~\cite{Welch1967} can be applied. The data length of each chunk $N_C $ is $N/n$ whereas the overall data length is $N$. We estimate the power spectral density $S(f)$ of the synthesized output $X$ referred to the beatnote frequency by
    \begin{equation}
    \begin{aligned}
        S(f) &=\frac{2C}{T}\sum_{k=1}^{2n-1}\left|\frac{X^{(k)}}{S_{1 \textrm{\_est}}S_{2 \textrm{\_est}}}\right|^2 \\
        &= \frac{2CT_S}{(2n-1)N_C} \sum_{k=1}^{2n-1} \left| \frac{y_1^{(k)}}{S_{1 \textrm{\_est}}}-\frac{y_2^{(k)}}{S_{2 \textrm{\_est}}} -  \delta \nu_{b \textrm{\_FR}}^{(k)} \right|^2,
    \end{aligned}
    \end{equation}
where superscript $(k)$ denotes the Fourier-transformed quantity using $k$-th chunk, $T$ is a measurement duration, $T_S$ is a sampling duration, and $C$ is a correction factor specific to the window function. In our case, the Hann window was applied and the correction $C=8/3$ was used.

\section{Demonstration of subtraction}\label{sec:4}
Two sets of data were recorded at different times. One is without the frequency noise excitation for estimating the noise floor as shown in Fig.~\ref{fig:beat_budget}. The other is with the frequency noise excitation enabled. In both cases, the frequencies of lasers were locked to the Fabry-Perot cavities all the time, producing a continuous train of time series data.

To obtain an accurate estimation of the sensing transfer function $S_j$, we experimentally measured the transfer functions between the beatnote and the error signals. The sensing transfer function was modeled by a single-pole system as
    \begin{equation}
        S_{j \textrm{\_est}}(f) = \frac{a_j \exp(-2\pi f \tau_j \mathrm{i})}{1 + \mathrm{i} f/f_{c,j}},
        \label{eq:TF}
    \end{equation}
where $a_j$, $\tau_j$ and $f_{c,j}$ are all constants of real values representing the optical gain in V/Hz, relative time delay in sec and cavity pole frequency in Hz, respectively. Since the free spectral range of the cavity is a sufficiently high value of $\sim$326~MHz comparing to the frequency band of interest, the single-pole model is accurate enough for our purpose. The measurement of the transfer functions includes also the response of the frequency readout $G_\textrm{FR}$. However, $G_\textrm{FR}$ was measured independently and the effect was removed by applying its inverse in the frequency domain.

\begin{figure}[btp]
\includegraphics[keepaspectratio, width=\linewidth]{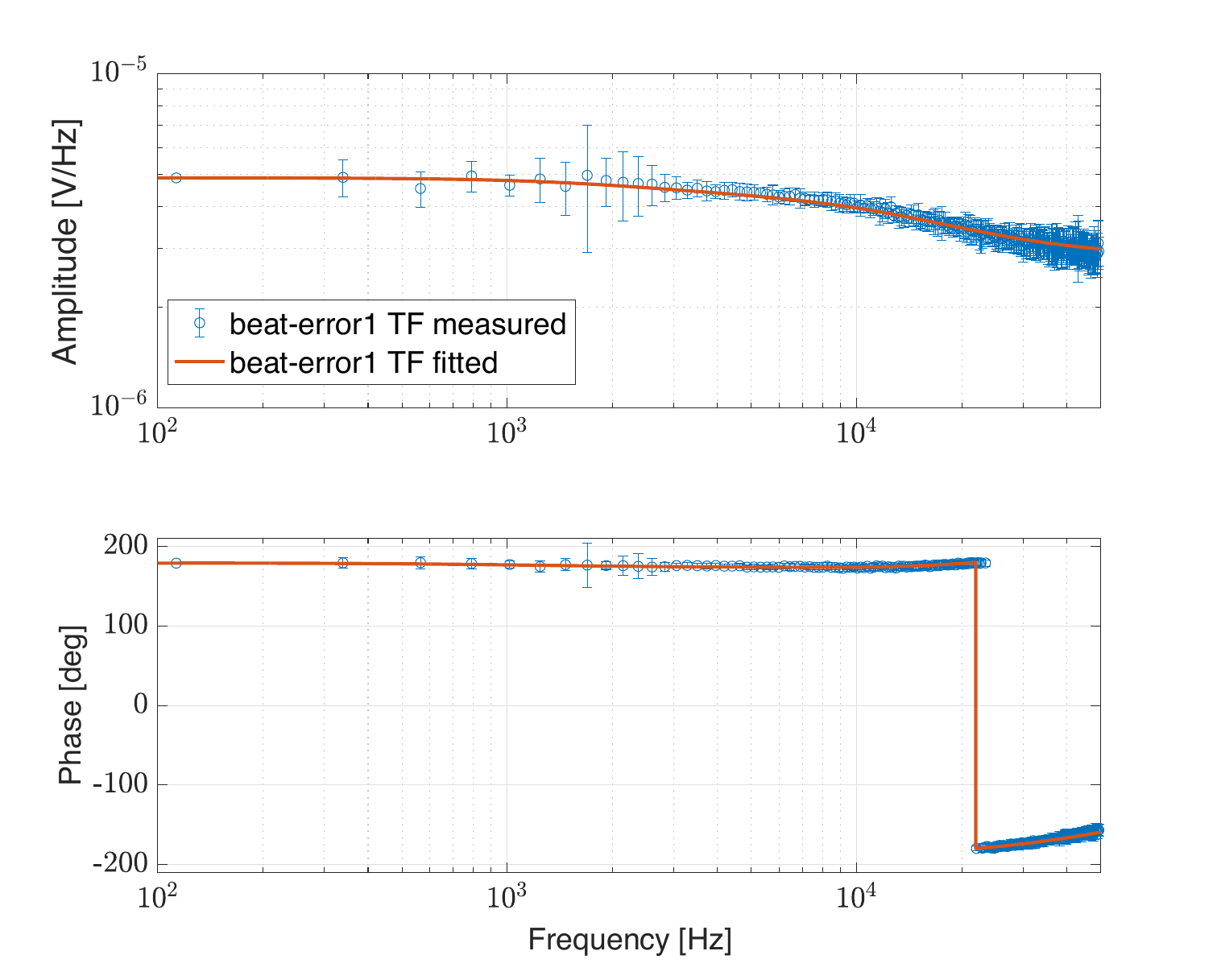}
\caption{\label{fig:be1TF} The measured transfer function between beat and error signal 1 (blue dots) and the fitted transfer function (orange solid). The dots indicate the values at the fundamental and harmonic frequencies of the excitation signal. The fit was performed using that portion of the signal. }
\end{figure}

The measurements of the sensing transfer functions were performed in-situ by utilizing the data set with the excitation enabled. The measured transfer function and the result of fitting to the model~(\ref{eq:TF}) are shown in Fig~\ref{fig:be1TF}. The models $S_{j \textrm{\_est}}$ with the fitted parameters are used in the subtraction.

\begin{figure*}[btp]
\includegraphics[keepaspectratio, width=\linewidth]{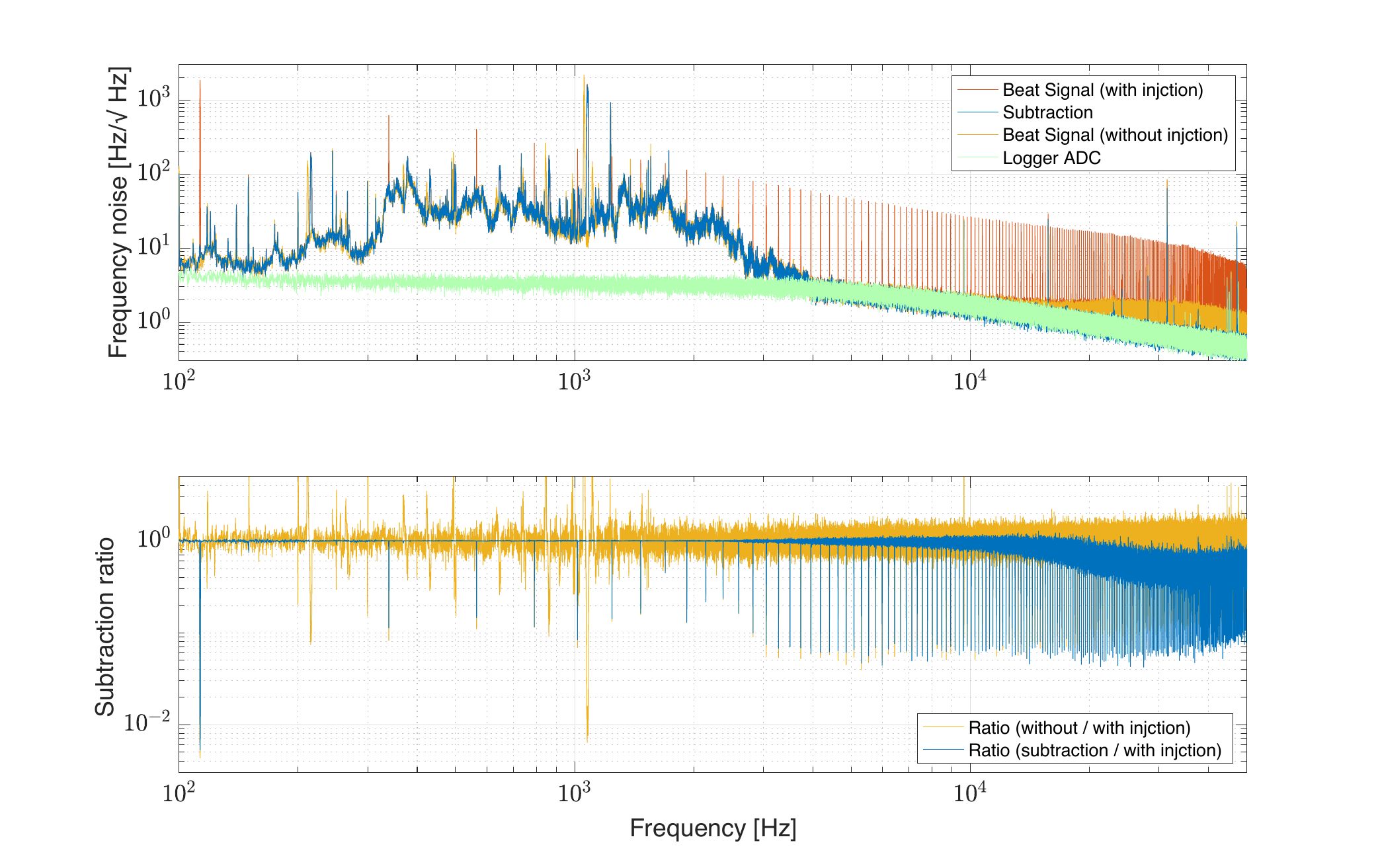}
\caption{\label{fig:subt1}The subtraction results. In the top panel, (Blue) spectrum with the offline subtraction applied. (Red) without the subtraction applied. (Yellow) when no excitation was applied. (Green) The ADC noise of the data logger for the beatnote frequency. In the bottom panel, the ratios of spectra are shown. (Blue) The ratio of the spectra with the subtraction applied over the one without excitation signals. Which corresponds to the inverse of the reduction ratio. (Yellow) The ratio of the spectra without the subtraction applied over the one without excitation signals.}
\end{figure*}

Fig.~\ref{fig:subt1} shows the amplitude spectral densities of the beatnote signals with and without the offline subtraction. The red curve was obtained while keeping the square wave excitation enabled. As expected, the higher order harmonics at frequencies of odd integer multiples of 113Hz are clearly visible in the spectrum. 

As seen in Fig.~\ref{fig:subt1}, the excitation peaks were successfully subtracted to the levels as low as the ambient noise floor not only in the low frequency region, where the sensing transfer functions $S_j$ are almost constant, but also at around the cavity pole and frequencies above. In particular, a reduction of a broad continuum is confirmed at frequencies above 10~kHz by a factor of two at most compared to the spectrum without the excitation. In this region, the laser frequency noises are expected to be the dominant components as summarized in Fig.~\ref{fig:beat_budget}. This means that, albeit limited, the subtraction has been achieved without the aid of a deliberate excitation in this frequency region.

\begin{figure*}[btp]
\includegraphics[keepaspectratio, width=\linewidth]{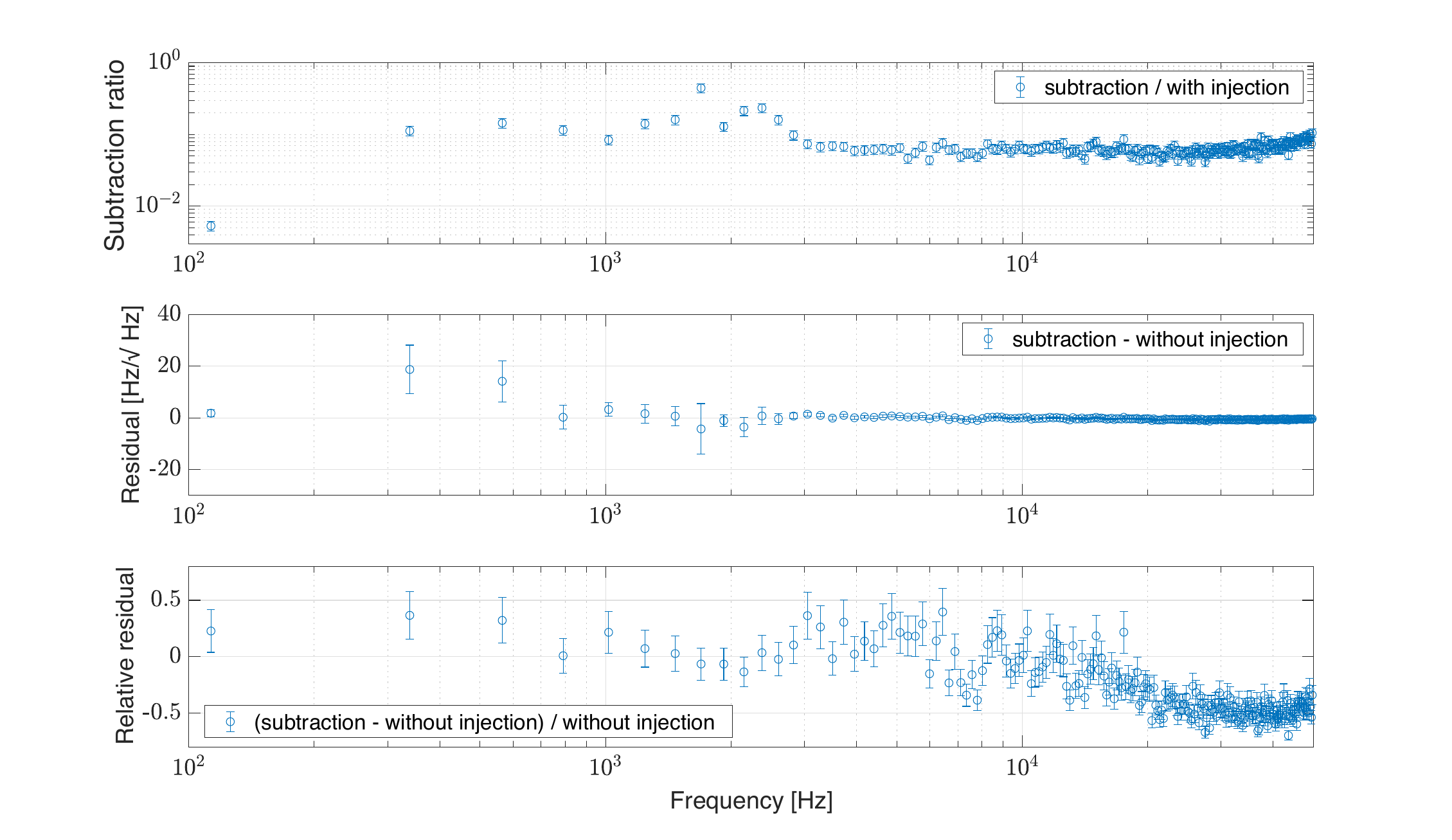}
\caption{\label{fig:residual}The top panel is the subtraction ratio at each injection peak. The middle panel is the amplitude noise residual for the without injection state of the after subtraction, and the bottom panel is the corresponding relative value. Error bars are derived from the 1-sigma range of the power spectrum estimation error.}
\end{figure*}

Figure~\ref{fig:residual} shows the subtraction ratio and the residuals at each excitation peak. The error bars in these plots are given by the 1-sigma uncertainties in the estimate of the power spectral density.

In the top panel of Fig.~\ref{fig:residual}, the best noise subtraction ratio of $0.53 \pm 0.08 ~\%$ is achieved at 113 Hz where the peak height is the highest relative to the ambient noise floor. The middle and bottom panels show respectively the residuals and the relative residuals normalized by the ambient noise floor. If the relative residual is 0, it means that the excitation is reduced to the noise floor. For example, some peaks, including the first to third and the fifth peaks at 113 Hz, 339 Hz, 565 Hz, and 1017 Hz, respectively, have significant residuals relative to the noise floor. On the other hand, in the high frequency region above 10 kHz, where frequency noise is expected to be predominant, not only the injection peaks but also the noise floor itself were reduced to the half of the original noise floor at best. 

\section{Discussion}\label{sec:5}
\subsection{Limiting factors in subtraction}

The residuals in the subtraction can be primarily caused by features that were not modeled in the transfer function implemented to the subtraction. At the same time, temporal variations in the system parameters could be critical for the peaks with fine subtraction ratios, such as the first peak. 

To consider these effects, $S_j$ is replaced to $S_j+\delta S_j$ by the stationary components $S_j$ and fluctuating components $\delta S_j$. Similarly, the estimated value $S_{j \textrm{\_est}}$ is replaced to $S_j+ \Delta S_j$ by the stationary component of the true value $S_j$ and the stationary deviation $\Delta S_j$. In addition, $\delta \nu_{\mathrm{b}}$ is replaced with $\delta \nu_{\mathrm{b}}+\delta n_{\mathrm{sen}}$ to account for the sensing noise $\delta n_{\mathrm{sen}}$.
Plugging these into Eq.~(\ref{eq:3}), the residual of frequency noise can be expressed as

\begin{equation}
\begin{aligned}
\frac{\Delta X_{\nu\_\mathrm{residual}}}{S_{1 \_ \text {est }} S_{2 \_ \text {est }}} & = \frac{1}{1+G_1} \frac{\Delta \mathrm{S}_1-\delta \mathrm{S}_1}{\mathrm{~S}_1+\Delta \mathrm{S}_1} \delta \nu_1 \\
& -\frac{1}{1+G_2} \frac{\Delta \mathrm{S}_2-\delta \mathrm{S}_2}{\mathrm{~S}_2+\Delta \mathrm{S}_2} \delta \nu_2+\delta n_{\mathrm{sen}}.\label{eq:resid}
\end{aligned}
\end{equation}
Here, we assumed $G_\textrm{FR \_est}=G_\textrm{FR}$ since the calibration with a known signal was performed, and the frequency readout response estimation is sufficiently accurate.

Moreover, As an effect not taken into account in the Eq.~(\ref{eq:resid}), the residuals may be worse due to the nonlinear effects as mentioned below at the frequencies which are integer multiple of the peaks with large subtraction.

We now discuss the limiting factors with special attention to the first to third and fifth peaks at 113 Hz, 339 Hz, 565 Hz, and 1017 Hz, respectively, which have significant residuals. The main limiting factors and these contributions for each peak are summarized in Table~\ref{table:1}. In what follows, we detail each factor.


\begin{table*}
\caption{\label{table:1}Residual noise at each peak and contribution of possible main factors.}
\begin{ruledtabular}
\begin{tabular}{llllll}
\begin{tabular}{l} Peak number 
\end{tabular} & 1st (113 Hz) & 2nd (339 Hz) & 3rd (565 Hz) & 5th (1017 Hz) & Above 5th \\
  \hline\hline
   \begin{tabular}{l} Subtraction ratio
\end{tabular} & $0.53\pm0.08 ~\%$
 & $11.3\pm1.7 ~\%$
 & $14.5\pm2.2 ~\%$
 & $8.4\pm1.3 ~\%$
 & $>4~\%$\\ \hline
    \begin{tabular}{l} Reduction ratio
\end{tabular} &$188\pm29$
&$8.9\pm1.4$
&$6.9\pm1.1$
&$11.9\pm1.8$
& $<25$\\ \hline
   \begin{tabular}{l} Noise residual 
\end{tabular} & \begin{tabular}{l} $1.8\pm1.4~\rm{Hz/ \sqrt{Hz}}$\\ ($0.23\pm0.19$)
\end{tabular} & \begin{tabular}{l} $18.8\pm9.4~\rm{Hz/ \sqrt{Hz}}$\\ ($0.37\pm0.21$)
\end{tabular} & \begin{tabular}{l} $14.2\pm7.9~\rm{Hz/ \sqrt{Hz}}$\\ ($0.32\pm0.20$)
\end{tabular} & \begin{tabular}{l} $3.3\pm2.6~\rm{Hz/ \sqrt{Hz}}$\\ ($0.22\pm0.19$)
\end{tabular} & \ \ \ \ -\\ \hline
   \begin{tabular}{l} Temporal variation\\ (typical value)
\end{tabular} & \multicolumn{4}{c}{$\sim0.81$~\%}\\ \hline
   \begin{tabular}{l} Noise residual\\ after tuning
\end{tabular} & \begin{tabular}{l} $1.8\pm1.4~\rm{Hz/ \sqrt{Hz}}$\\ ($0.23\pm0.19$)
\end{tabular} & \begin{tabular}{l} $18.3\pm9.4~\rm{Hz/ \sqrt{Hz}}$\\ ($0.36\pm0.21$)
\end{tabular} & \begin{tabular}{l} $6.2\pm7.3~\rm{Hz/ \sqrt{Hz}}$\\ ($0.14\pm0.18$)
\end{tabular} & \begin{tabular}{l} $1.7\pm2.4~\rm{Hz/ \sqrt{Hz}}$\\ ($0.11\pm0.17$)
\end{tabular} & \ \ \ \ -\\ \hline
   \begin{tabular}{l} Nonlinear effects\\ (estimated maximum)
\end{tabular} & \ \ \ \ \ \ \ - &\  $ 8.3\pm0.9~\rm{Hz/ \sqrt{Hz}}$ & \ \ \ \ \ \ \ - & \ $2.8\pm0.2~\rm{Hz/ \sqrt{Hz}}$ & \ \ \ \ -\\  \hline
 \begin{tabular}{c} Sensing noises
\end{tabular} & \multicolumn{4}{c}{Not effective in the demonstration with the noise injection}\\
\end{tabular}
\end{ruledtabular}
\end{table*}

\vskip\baselineskip
\noindent
\textbf{Unmodeled features in transfer function}

The sensing transfer function $S_j$ was estimated by fitting the measurement as shown in Fig~\ref{fig:be1TF}. To confirm the accuracy of the transfer function model, the phase and amplitude of the transfer function were tuned to minimize the residuals at the peak frequencies where there had been residuals. If there is a large improvement in the noise residual, it could indicate that there exist some features such as peaks and notches that are not taken into account in the model and implies that $\Delta S_j$ in Eq.~(\ref{eq:resid}) is significant. It is relatively difficult to find such unmodeled features in the regions where the signal-to-noise ratio of the measurement is low. 

The residual noises after tuning are shown in the row 6 of Table~\ref{table:1}. There was no significant improvement in the fundamental peak at 113 Hz and the second peak at 339 Hz. On the other hand, the residuals improved by approximately half for the third peak at 565 Hz and the fifth peak at 1017 Hz. Therefore, it is possible that a local feature in the transfer function is the main cause of the deterioration in the subtraction ratio for the third and fifth peaks whereas the residuals in the first and second peaks seem to be due to other factors.

\vskip\baselineskip
\noindent
\textbf{Temporal variation in system}

Temporal variations in the system parameters represented by $\delta S_j$ in Eq.~(\ref{eq:resid}) alter the sensing transfer function $S_j$ as a function of time. This could limit the subtraction ratio, too. To study the stability of the parameters associated with the sensing transfer function, the transmitted light of the cavities have been monitored.

The RMS (root mean square) of the RIN (Relative Intensity Noise) was measured to be 0.81 \% for cavity 1. This transmitted light RIN could be attributed to the variation of the incident laser power into the cavity or the variation of the optical gain. The latter is caused by changes in the transmittances of the cavity mirrors or the internal loss of the cavity. However, it was difficult to distinguish which parameter dominated in the RIN. Therefore, we consider the case where one of them is dominant.

The amplitude variation of the PDH signal can be directly regarded as the variation of the sensing transfer function amplitude $ \delta a_j$. Therefore, we evaluated $\delta a_j$ as a function of the RIN for the transmitted light. For convenience, we introduce the relation $\delta a_j/ a_j = \alpha RIN$ with $\alpha$ being a constant of real value. We show that the coefficient $\alpha$ takes different values depending on what parameter is the cause for the variations. 

In the case where the cause of the variation is either the power of the incident light, the transmittance of the cavity input mirror, or the loss in the cavity, the coefficient becomes $\alpha=1$. These parameters affect the PDH signal and the transmitted light intensity in the same way. On the other hand, in the case where the variation of the PDH signal is caused by the variation in the transmittance of the end mirror, the coefficient takes a value much smaller than unity or $\alpha \ll1$. This is confirmed by performing the partial differentiation of the expression for the PDH signal and the transmitted light with respect to the end mirror transmittance. We find that the coefficient can be approximated to be $T_e/L$ where $L$ is the loss in the cavity and $T_e $ is the power transmittance of the end mirror. Assuming $T_e = \  \sim 10^{-5}$ and $L= \ \sim 10^{-2}$ for our experimental setup, we estimated the coefficient $\alpha$ to be on the order of $ 10^{-3}$. In this case, the PDH signal is almost unaffected even when the transmitted light power fluctuates.

Since sub-percent levels of variations are unlikely to occur in the mirror transmittance, the variations may be mainly due to the incident laser power and the loss in the cavity. Thus, the relative variations in the PDH signal should equal the transmitted light RIN or $\alpha=1$.

Although the RMS value of 0.81\ \% for RIN represents the typical value of variability, this measurement suggests that the amplitude of the sensing transfer function $a_1$ fluctuates at the sub-percent level. Given the fact that the subtraction ratio at the first injection peak at 113 Hz was about 0.5\ \% in accuracy, time variations of $a_1$ may be the main factor for the residuals. Further investigation is necessary to achieve a more stable system in the future.

A similar RIN of about 1.1\ \% is also observed for cavity 2. However the influence of laser 2 on the injection peaks is negligible and does not affect the subtraction ratio.

For the peaks at the second harmonic and higher, the subtraction accuracies are about 4\ \% at best. Thus, the effect of temporal variation is not critical.

\vskip\baselineskip
\noindent
\textbf{Sensing noises}

Noise coupled after the laser light is split into the back-link and cavity sides becomes an uncorrelated component between the beatnote and error signals. Therefore, such noises, expressed as $\delta n_{\mathrm{sen}}$ in Eq.~(\ref{eq:resid}), cannot be subtracted and contaminate the detector sensitivity directly. The main candidates for such sensing noises are phase noise introduced by mirror reflections, optical fibers, air disturbances in the free space region, etc., or noise introduced from the phasemeter, and ultimately shot noise in the backlink heterodyne.

As shown in Fig~\ref{fig:beat_budget}, the unidentified noise below 4~kHz and ADC noise of the data logger at high frequencies are examples. Although these noises were not a problem in our demonstration because of the relatively high excitation peaks, they would be critical factors if a lower noise level is pursued. Even if an excitation is applied to the laser frequencies, the sensing noise must be sufficiently reduced because the excitation with a too high amplitude would worsen the state of resonance and lead to an enhancement of the nonlinear effects.

For phase noise, there are ideas for common-mode rejection of fiber phase noise by acquiring beat signals at two photodetectors, as shown in Fig~\ref{fig:BLFP2}. An alternative is to lock the length of the back-link heterodyne path to some stabilized reference. Additionally, it is essential to develop a low noise phasemeter. Shot noise can be reduced without reducing the light power sent to the cavity using the heterodyne squeezing scheme~\cite{Anai2023}.

\vskip\baselineskip
\noindent
\textbf{Nonlinearities}

The PDH signal is known to come with nonlinearity. The response is not a mere linear line but exhibits an S-shaped curve with respect to the laser frequency. This directly means that the error signal contain higher order terms together with the linear. When the subtraction is performed for a specific frequency, noises are expected to worsen at the multiplier frequency. 

\begin{figure}[btp]
\includegraphics[keepaspectratio, width=\linewidth]{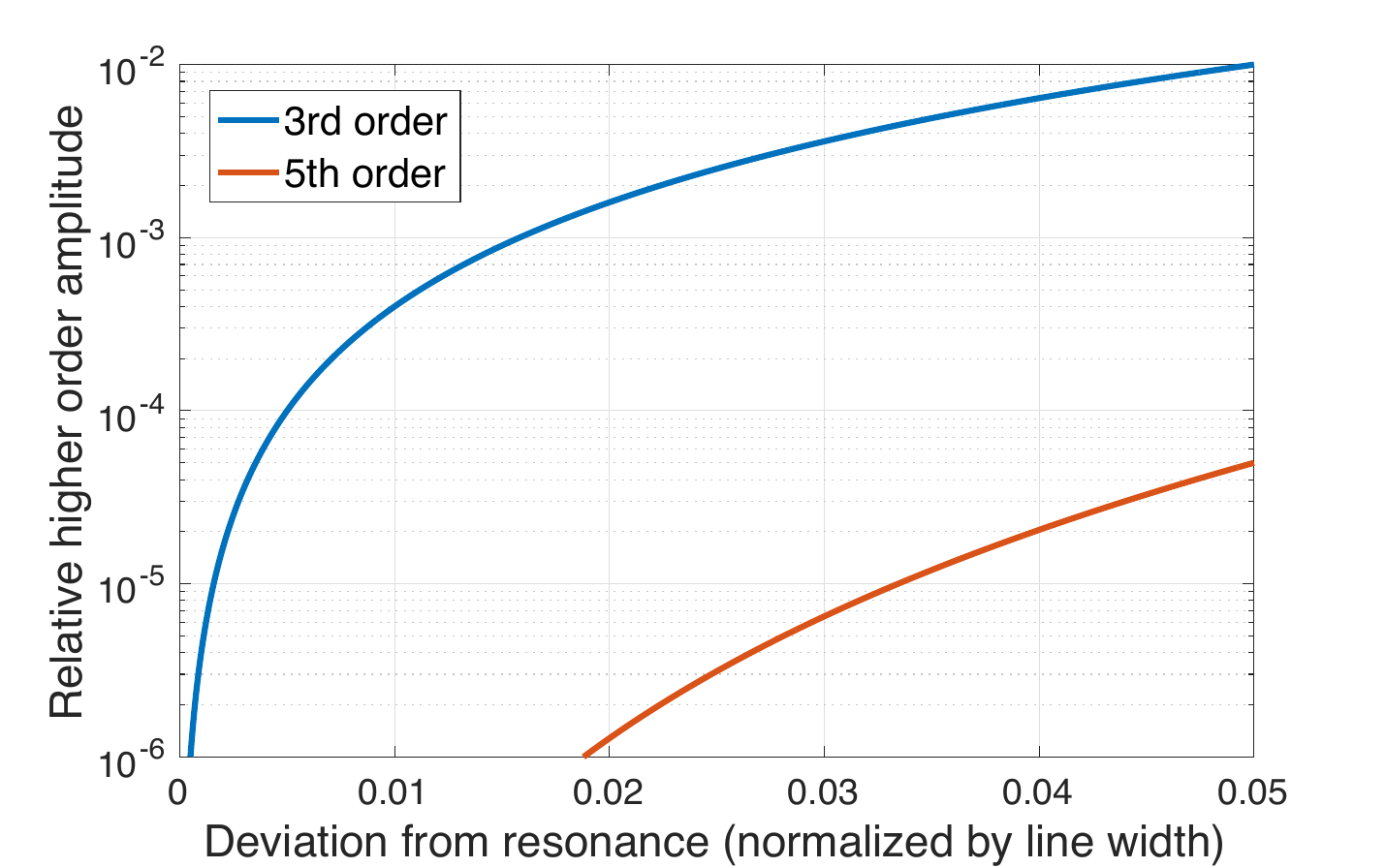}
\caption{\label{fig:PDH_HO}The contribution of the higher-order terms relative to the first-order term in the PDH signal. The horizontal axis is the deviation from resonance normalized by the cavity linewidth.}
\end{figure}

The PDH signal $V_{\textrm{PDH}}$ can be expressed as follows, assuming that the modulation sidebands are sufficiently far apart from the resonant frequencies of the cavity.

\begin{equation}
V_{\textrm{PDH}} \propto \frac{\sin \phi}{1+R^2-2 R \cos \phi}, \label{eq:8}
\end{equation}
where $\phi$ is the phase deviation from a resonance point and $R$ is the product of the amplitude reflectivities of the input and output mirrors of the cavity. $R$ is given as follows using the cavity finesse $\mathcal{F}$.

\begin{equation}
R=\frac{2 \mathcal{F}^2+\pi^2-\pi \sqrt{4 \mathcal{F}^2+\pi^2}}{2 \mathcal{F}^2}.
\end{equation}
Figure~\ref{fig:PDH_HO} shows the values of the higher-order components normalized by the linear-order component when Eq.~(\ref{eq:8}) is expanded by the power series of $\phi$. We have assumed that the operating point is at a perfect resonance without considering the effect of offsets for simplicity. Note that this effect would be larger and the even-order contents would appear in the presence of offsets.

The measurement of the transmitted light suggested that the laser frequency fluctuated around the resonance by 2~kHz at most during the excitation test. Therefore, we assume a deviation of about 1~kHz from the resonance. This corresponds to approximately $1/30$ of the cavity linewidth. In this case, the effect of the third-order term, which is the most dominant among the higher-order terms, is $\sim4.5 \times 10^{-3} $ for the first-order term.
Therefore, the excitation can introduce the additional noise of the peak amplitude multiplied by this factor at a tripled frequency via the nonlinear effect. The effect of the fundamental peak at 113 Hz on the second peak at 339 Hz and the effect of the second peak at 339 Hz on the fifth peak at 1017 Hz are shown in row 7 of Table~\ref{table:1}.
At the second peak, this effect could be the main cause of the residual. Also, part of the residuals in the fifth peak can be caused by this effect.
For harmonic peaks higher than the fifth order, the peak height monotonically decreases by the nature of the square waves. Therefore, the nonlinear effects are less likely to appear.

\vskip\baselineskip
\noindent
\subsection{Implications for the space gravitational wave antenna and future prospect}
Assuming $10^{-23} \ /\rm{\sqrt{Hz}}$ as the target sensitivity for the amplitude of gravitational waves, a noise level of $6 \times 10^{-10} \ \rm{Hz/\sqrt{Hz}}$ will be required for the the laser frequency of $6 \times 10^{14} \ \rm{Hz}$ with a signal-to-noise ratio of 10 incorporated.
Such a noise levels have to be achieved by allocating the capabilities of the initial stability of the laser frequency, the suppression gain of the frequency-locked loop, and the offline noise subtraction, respectively. The reduction by a factor of $10^8$ is necessary even if we assume an excellent initial stability of $0.1 \ \rm{Hz/\sqrt{Hz}}$ which is currently achieved on a laboratory scale~\cite{Numata2004}. 

The proposal paper~\cite{BLFP} assumed a conservative value of $10^4$ for the suppression factor in the frequency-locked loop. The rest of the reduction by $10^4$ must be achieved by the offline noise subtraction. In this scenario, the reduction ratio of $\sim$200 achieved in our experiment needs to be improved by another factor of 50.

A more aggressive control gain in the frequency-locked loop might achieve the requirement even with the current reduction ratio of the subtraction. However, in future missions such as DECIGO will require $10^{-24} \ /\rm{\sqrt{Hz}}$ or better in the sensitivity~\cite{DECIGO}. Moreover, regardless of the scenario, the relaxation of design requirements for hardware, such as lasers and control systems, resulting from the improvement of the reduction ratio of the subtraction, is important for spacecraft.

For these reasons, we are planning to upgrade the experimental setup to achieve a high reduction ratio of the subtraction over $10^4$. To achieve this, we have to identify the dominant factors causing time variations and stabilize the system. It is also important to address nonlinear effects to achieve high reduction ratio at broad frequencies.

Possible hardware upgrades include the introduction of a vacuum environment, a vibration isolation system, and a monolithic optical system. We also plan to introduce a much tighter frequency-locked loop to reduce nonlinear effects, as well as an additional system controlling the offset in the operating point using the transmitted light.

As for improvements to the subtraction process, we plan to implement an adaptive transfer function that takes into account the temporal variation of the system~\cite{Darkhan2016}, an advanced subtraction process that takes into account nonlinear effects, and a time domain process to avoid the undesirable residuals expected in the finite-length Fourier transform. 
\section{conclusion}\label{sec:6}
We have experimentally demonstrated for the first time the laser frequency noise subtraction in the back-linked Fabry-Perot interferometer. A miniature of the interferometer was built on a tabletop under atmospheric pressure where two identical Fabry-Perot cavities simulate a single set of the back-linked Fabry-Perot interferometer. Deliberately increasing noise in the laser frequency, we succeed in subtracting laser frequency noises in the broad frequency band, including the cavity pole frequency. We confirmed that the highest reduction ratio reached $188 \pm 29$ at 113~Hz. This marks the experimental proof of the concept for the noise subtraction for the first time.

Currently, the subtraction performance at the frequency where the highest reduction ratio has been achieved seems to be limited by temporal variations of the system parameters. In addition, a large noise amplitude at a frequency may exacerbate residuals at frequencies integer multiple to the original through the nonlinearity of the error signal. Such nonlinear effects could become a critical issue when performing large subtractions over a broad frequency bandwidth in future. To demonstrate a higher reduction ratio, we plan to upgrade the setup by stabilizing the experimental system and implementing an advanced subtraction process.
\begin{acknowledgments}
This work was supported by JSPS KAKENHI Grant Number JP20H01938 and Research Grants in the Natural Sciences by the Mitsubishi foundation \#2 in Japan fiscal year 2020. KI thanks the KAGRA collaboration for an electronic design software. The Fabry-Perot cavity spacer and other optical components were machined by the Advanced Machining Technology Group in the Institute of Space and Astronautical Science, JAXA.
\end{acknowledgments}

\nocite{*}

\bibliography{main} 


\end{document}